\documentclass[showpacs,amsmath,amssymb,twocolumn,superscriptaddress]{revtex4}
\usepackage{graphicx,float}
\usepackage[all]{xy}
\usepackage{amsmath}
\usepackage{amssymb}
\usepackage{color}

\newcommand{\bb}{\bibitem}
\newcommand{\bes}{\begin{subequations}}
\newcommand{\ees}{\end{subequations}}
\def\ben{\begin{eqnarray}}
\def\een{\end{eqnarray}}
\def\be{\begin{equation}}
\def\ee{\end{equation}}
\def\sech{\text{sech}}
\def\sech{\textrm{sech}}

\begin{document}
\title{A Note on Asymmetric Thick Branes}
\author{D. Bazeia}
\email{bazeia@fisica.ufpb.br}
\affiliation{Departamento de F\'\i sica, Universidade Federal da Para\'\i ba, 58051-970 Jo\~ao Pessoa, PB, Brazil}
\affiliation{Departamento de F\'\i sica, Universidade Federal de Campina Grande, 58109-970 Campina Grande, PB, Brazil}
\author{R. Menezes}
\email{rmenenezes@dce.ufpb.br}
\affiliation{Departamento de F\'\i sica, Universidade Federal de Campina Grande, 58109-970 Campina Grande, PB, Brazil}
\affiliation{Departamento de Ci\^encias Exatas, Universidade Federal da Para\'\i ba, 58297-000 Rio Tinto, PB, Brazil.}
\author{R. da Rocha}
\email{roldao.rocha@ufabc.edu.br}
\affiliation{CMCC,
Universidade Federal do ABC, 09210-170, Santo Andr\'e, SP, Brazil.}

\date{\today}
\begin{abstract}

In this work we study asymmetric thick braneworld scenarios, generated after adding a constant to the superpotential associated to the scalar field. 
We study in particular models with odd and even polynomial superpotentials, and we show that asymmetric brane can be generated irrespective of
the potential being symmetric or asymmetric.  We study in addition the nonpolynomial sine-Gordon-like model, also constructed with the inclusion of a constant in the 
standard superpotential, and we investigate gravitational stability of the asymmetric brane. The results suggest robustness of the new braneworld
scenarios and add further possibilities for the construction of asymmetric branes.
\end{abstract}

\maketitle

%%%%%%%%%%%%%%%%%%%%%%%%%%%
\section{Introduction}

The concept of braneworld with a single extra dimension of infinite extent \cite{RS} plays a fundamental role in high energy physics and cosmology. In such braneworld scenario, the particles of the standard model are confined to the brane, while the graviton can propagate in the whole space \cite{RS,GW,F,C}; see also \cite{G} for further details.   

Most braneworld scenarios assume a $\mathbb{Z}_2$-symmetric brane, as motivated by the Horava-Witten model \cite{howi}. Notwithstanding, more general models that are not directly derived from M-theory, are obtained by relaxing the mirror symmetry across the brane \cite{kraus,stoica,gergelyfri,bowcock,carter,koyama11,dolozel,ida,perkins,appleby, gergelymaar,shtanov,ocalla,nozari,charmousis,koyapad, padilla,guerrero,zhang}. Asymmetric  branes are usually considered in the literature as a framework to models where the $\mathbb{Z}_2$ symmetry 
is not required. The parameters in the bulk, such as the gravitational and cosmological constants, can differ on either sides of the brane.
The term coined ``asymmetric'' branes means also that the gravitational parameters of the theory can differ on either side of the brane. 
In some cases, the Planck scale on the brane differs on either side of the domain wall \cite{padilla,grepa}. One of the  prominent features regarding  asymmetric braneworld models is that they present some self-accelerating solutions
without the necessity to consider an induced gravity term in the brane action \cite{nozari}.

Braneworld models without mirror symmetry have been investigated in further  aspects. For instance,  different black hole masses  were obtained on the two sides on the brane \cite{kraus}, as well as different cosmological constants on the two sides  \cite{dolozel,perkins,bowcock,
carter,stoica,gergelyfri}. In addition, one side of the brane can be unstable, and the other one stable \cite{shtanov,koyapad}. If the $\mathbb{Z}_2$  symmetry is taken apart, it is also possible to get infrared modifications of gravity, comprehensively considered in the asymmetric models in \cite{padilla} and in the hybrid asymmetric DGP model in Ref. \cite{charmousis} as well. In both models, the  Planck mass and the cosmological constant  are different at each side of the brane. In the asymmetric model, there is a hierarchy between the Planck masses and the $AdS$ curvature scales on each side of the brane \cite{ocalla}. 

Asymmetric braneworlds can be further described by thick domain walls in a geometry that  asymptotically is led 
to different cosmological constants, being de Sitter $(dS)$ in one side and anti de Sitter $(AdS)$ in the other one, along the perpendicular direction to the wall. The  asymmetry regarding the braneworld arises as the scalar field interpolates between two non-degenerate minima of a scalar potential without $\mathbb{Z}_2$ symmetry.  Asymmetric static double-braneworlds with two different walls were also considered in this context, embedded in a $AdS_5$ bulk. 
Furthermore, an independent derivation of the Friedmann equation was presented in the simplest for an $AdS_5$ bulk, with different cosmological constants on the two sides of the brane \cite{gergelymaar}. Finally, a braneworld that acts as a domain wall between two different five-dimensional bulks was considered, as a solution to Gauss-Bonnet gravity \cite{dolozel,nozari}. Models with moving branes  have been  further considered \cite{kkkk} in order to break the reflection symmetry of the brane model \cite{battye,carter,uzan,kraus,stoica,perkins,klkl}.

In the current work, we shall further study asymmetric branes, focusing on general features, which we believe can be used to better qualify the braneworld scenario as symmetric or asymmetric.
The key issue is to add a constant to the superpotential that describe the scalar field, which affects all the braneworld results. In particular, one notes that it explicitly alters the quantum potential, that responds for stability, evincing and unraveling prominent physical features, as the asymmetric localization of the graviton zero mode.
We analyze all the above mentioned features of asymmetric branes in the context of superpotentials containing odd and even power in the scalar field, up to the third and fourth order power, respectively, and the sine-Gordon model.

The investigation starts with a brief revision of the equations that govern the braneworld scenario under investigation, getting to the first-order framework, where first-order differential equations solve the Einstein and field equations if the potential has a very specific form. We deal with asymmetric branes, in the scenario where the brane is generated from a kink of two distinct and well-known models, with the superpotentials being odd and even, respectively. Subsequently, the sine-Gordon model is also studied and the associated linear stability is investigated. The quantum potential and the graviton asymmetric zero mode are explicitly constructed.

%%%%%%%%%%%%%%%%%%%%%%%%%%%%
\section{The framework}
We start with a five-dimensional action in which gravity is coupled to a scalar field in the form
\be
{\cal S}=\int d^5 x\sqrt{|g|} \left(-\frac14{R}+{\cal L} \right),
\ee
where 
\be
{\cal L}=\frac12 \nabla_a \phi \nabla^a\phi - V(\phi),
\ee
and $a=0,1,...,4$. We are using $4\pi G=1$, with field and space and time variables dimensionless, for simplicity. By denoting $V_\phi = dV/d\phi$, the Einstein equations and the Euler-Lagrange equation are $G_{ab}=T_{ab}$ and $\nabla^a \nabla_a \phi+ V_\phi=0$, respectively. We write the metric as
\be
ds^2=e^{2A}\eta_{\mu\nu} dx^\mu dx^\nu - dy^2,
\ee
where $A=A(y)$ describes the warp function and only depends on the extra dimension $y$. Taking the scalar field with same dependence, we obtain
\bes\label{EoM}
\ben
A^{\prime\prime}+\frac23 \phi^{\prime2}&=&0, \\
A^{\prime2}- \frac16 \phi^{\prime2}  + \frac16 V(\phi)&=&0,\\
\phi^{\prime\prime}+4 \phi^\prime A^\prime-V_\phi&=&0,
\een
where prime stands for derivative with respect to the extra dimension.
\ees
As one knows, solutions to the first-order equations 
\bes
\ben
\phi^\prime &=& \frac12 W_\phi,\label{foA}\\ 
A^\prime&=&-\frac13 W(\phi), \label{firstorderA}
\een\ees
also solve the equations \eqref{EoM}, if the potential has the following form
\be
V=\frac18 W_\phi^2 - \frac13 W^2\label{potential}
\ee 
where $W=W(\phi)$ is a function of the scalar field $\phi$. This result shows that if one adds a constant to $W$, it will modify
the potential, and consequently, all the results that follow from the model.

To better explore this idea, we study three distinct models, which generalize previous investigations.

%%%%%%%%%%%%%%%%%%%%%%%%
\subsection{The case of an odd superpotential}\label{sec:2a}

Let us introduce the function
\be
W_c(\phi)=-\frac23 \phi^3+2\phi  +c\,.
\ee
where $c$ is a real constant. This $W$ is an odd fucntion for $c=0$. The potential given by Eq. \eqref{potential} has now the form

\be\label{p4V}
V(\phi)=\frac12 \left(1-
\phi^2\right)^2-\frac{1}{3}\left( {c}+2\phi -\frac23 \phi^3\right)^2\,.
\ee
It has the $\mathbb{Z}_2$ symmetry only when $c=0$; the symmetry is now $\phi \mapsto -\phi$ and $c\mapsto -c$. There are minima at $\phi_\pm =\pm 1$, with
\be
W_c(\phi_\pm)=\pm\frac43+c\,.
\ee
 The maxima depend on $c$ and are  solutions of the algebraic equation $4\phi_{max}^3 -{39}\phi_{max}-2c=0$. See Fig.~\ref{fig1}.
%%%%%%%%%%%%%%%%%%%%%%%%%%%%%%%%%%
\begin{figure}[h] 
\includegraphics[scale=0.45]{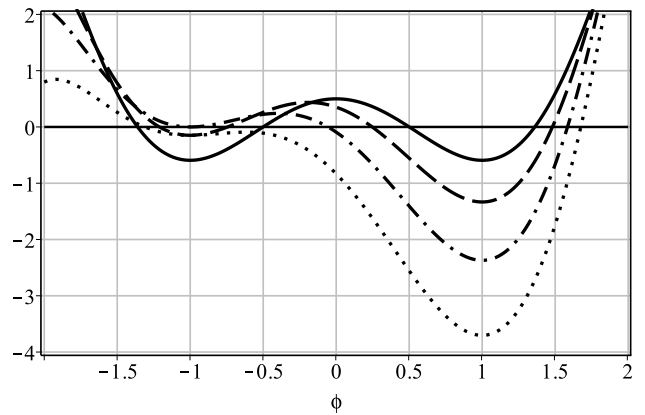} 
\caption{\label{fig1}The potential given by Eq. \eqref{p4V} for $c=0$ (black line), $c=2/3$ (dashed curve), $c=4/3$ (dot-dashed line), and $c=2$ (dotted line).}
\label{fig1}
\end{figure} 
%%%%%%%%%%%%%%%%%%%%%%%%%%%%%%%%%%

We focus attention on the scalar field. There is a topological sector connecting the minima. In this sector, there are two solutions (kink and antikink), which can be mapped to one another when $\phi \mapsto - \phi$ or $y \mapsto - y$. Thus, we  study solely the kinklike solution.
The first order equation for the field does not depend on the parameter $c$, and is provided by
$\phi^\prime=1-\phi^2$. The solution for this equation is
\be
\phi(y)=\tanh(y),\label{p4solution}
\ee 
with $\phi(\pm \infty)=\phi_\pm=\pm1$. It obeys $\phi(y)=-\phi(-y)$, and it connects symmetric minima. 

The warp function can be obtained by using Eq.~\eqref{firstorderA}
\be\label{p4warp}
A(y)=-\frac1{9}\tanh^2 \left( y \right) +\frac49\ln  \left(  \sech\left( y \right)  \right) - \frac{c}{3}\,y\,,
\ee
with $A(0)=0$. We note that the behavior of this function outside the brane can be written as
\bes
\ben
A_{\pm\infty}(y)\!&=&\!-\frac13 W(\phi_\pm ) |y|\! = \!- \frac13\left(\frac43\pm{c}\right) \!\!|y|
\een
and asymptotically the five-dimensional cosmological constant is 
\ben
\Lambda_{5_\pm}\equiv V(\phi({\pm\infty}))\!&=\!&-\frac13\left(\frac4{3}\pm {c}\right)^2\,.
\een
\ees
If $c=0$, the brane is symmetric, connecting two  regions in the $AdS_5$ bulk, with the same cosmological constant $\Lambda_{5_\pm}=-16/27$.  If $|c|>4/3$,  the warp factor $e^{2A}$ diverges. In all other cases, the warp factor describes asymmetric braneworlds. If $|c|=4/3$, the brane connects an $AdS_5$ bulk and a five-dimensional Minkowski ($\mathbb{M}_5$) bulk, with $\Lambda_{5+}=-64/27$ [$\Lambda_{5+}=0$] and $\Lambda_{5-}=0$ [$\Lambda_{5-}=-64/27$], when $c=4/3$ [$c=-4/3$]. If $0<|c|<4/3$, the brane connects two distinct $AdS_5$ bulk spaces. 

 In Fig.~\ref{fig2} we plot the profile of the warp factor, for some values of $c$.  The solid curve represents the symmetric profile (Brane I). The dashed curve represents the asymmetric brane that connects two $AdS_5$ bulk spaces,  with $c=2/3$, $\Lambda_{5+}=-4/3$, and $\Lambda_{5-}=-4/27$ (Brane II). The dot-dashed curve represents the case where the brane connects an $AdS_5$ to a $\mathbb{M}_5$ bulk (Brane III). Finally, the dotted curve represents the divergent warp factor with $c=2$. 
   
%%%%%%%%%%%%%%%%%%%%%%%%%%%%%%%%%%
\begin{figure}[h] 
\includegraphics[scale=0.45]{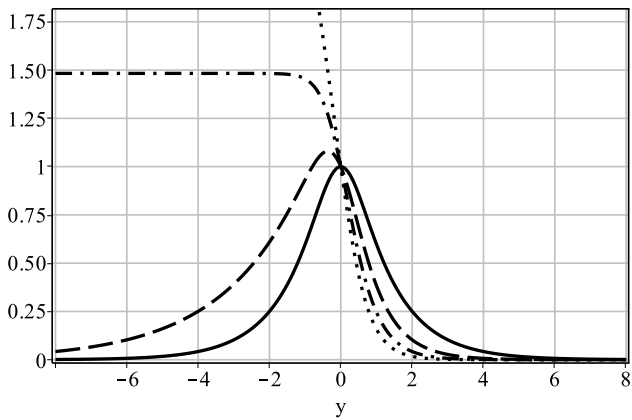} 
\caption{\label{fig2}The warp factor $e^{2A}$ with $A$ given by Eq. \eqref{p4warp} for $c=0$ (solid line), $c=2/3$ (dashed line), $c=4/3$ (dot-dashed line), $c=2$ (dotted line). We fix $A(0)=0$.}
\label{fig2}
\end{figure} 
%%%%%%%%%%%%%%%%%%%%%%%%%%%%%%%%%%%

\begin{figure}[h] 
\includegraphics[scale=0.45]{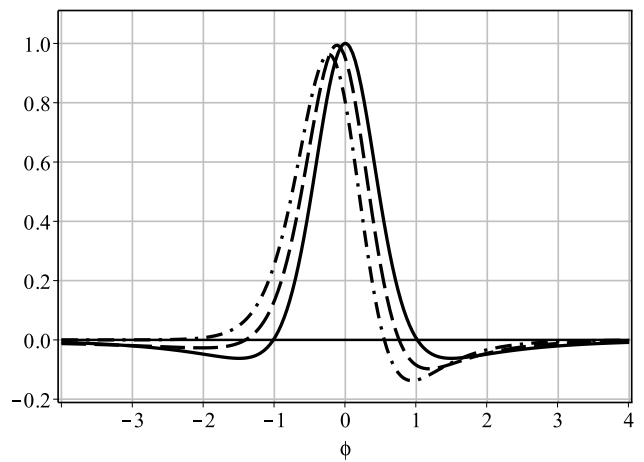} 
\caption{The energy density given by Eq.\eqref{rhop4} for $c=0$ (solid line), $c=2/3$ (dashed line) and $c=4/3$ (dot-dashed line).}
\label{fig3}
\end{figure}

The energy density is given by
\be
\rho(y)=-e^{2A} {\cal L},
\ee
and it also depends on $c$
\ben\label{rhop4}
\rho(y)&=&e^{2A} \Big( \nonumber
{c^2}-\frac{16}{81}+\frac4{27}\sech^4(y)+
\frac{31}{81}\sech^6(y)\\&& -\frac{4c}{27}\tanh(y) (2+\sech^2(y)).
\Big)
\een
It is symmetric only for $c=0$. When $c\neq0$,  there exist contributions to the asymmetry from both the warp factor $e^{2A}$ and the Lagrange density ${\cal L}$ as well. In Fig.~\ref{fig3}, we depict the profile for Brane I (solid curve), II (dashed curve) and III (dot-dashed curve). Therefore, for $c\neq0$, the warp factor and the energy density are asymmetric.

This model shows that although the kinklike solution connects symmetric minima, the potential is asymmetric and gives rise to braneworld scenario which is also asymmetric, unless $c=0$.

%%%%%%%%%%%%%%%%%%%%%%%%%%%%%%
%%%%%%%%%%%%%%%%%%%%%%%%%%%%%%

%%%%%%%%%%%%%%%%%%%%%%%%%%%%%%%%%%%%%%%%%%%%%%%%%
\subsection{The case of an even superpotential}

Let us now investigate a different model, which presents kinklike solution that connects asymmetric minima. It is defined by
\be
W_c(\phi)=-\frac12 \phi^4 + \phi^2 + c\,.
\ee
This $W$ is an even function of $\phi$. The potential given by Eq. \eqref{potential} is now
\be\label{p6V}
V(\phi)=\frac12 \phi^2 (1-\phi^2)^2 - \frac13 \left({c}+ \phi^2-\frac12 \phi^4  
 \right)^2\,.
\ee
It is depicted in Fig.~\ref{fig4}, and it has the $\mathbb{Z}_2$ symmetry for any value of the parameter $c$. The minima are  $\phi_-=-1$, $\phi_0=0$ and $\phi_+=1$, with 
\be
W_c(\phi_\pm)=\frac12+c\,\,\,\,\,\,{\rm and}\,\,\,\,\,\,W_c(\phi_0)=c
\ee

%%%%%%%%%%%%%%%%%%%%%%%%%%%%%%%%%
\begin{figure}[h!] 
\includegraphics[scale=0.45]{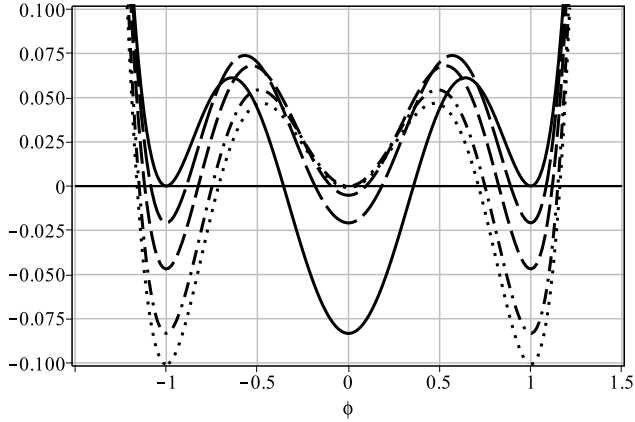} 
\caption{\label{fig4}The potential given by Eq. \eqref{p6V} for $c=-1/2$ (solid line), $c=-1/4$ (long dashed line), $c=-1/8$ (dashed line), $c=0$ (dot-dashed line), $c=1/20$ (dotted line).}
\label{fig4}
\end{figure} 
%%%%%%%%%%%%%%%%%%%%%%

There are two topological sectors. The first connects the minima $\phi_-$ and $\phi_0$, while the second connects the minima $\phi_0$ and $\phi_+$. Note that these sectors can be mapped by taking the transformation $\phi \mapsto -\phi$. In each sector, there are two solutions (kink and antikink); they can be mapped into each other with the coordinate transformation $y \mapsto -y$. Thus, we only study one of these solutions. Using the first-order equation, $\phi^\prime=\phi \left(1-\phi^2\right)$, we find the solution 
\be
\phi(y)=\frac{\sqrt{2}}{2} \sqrt{1+\tanh(y)},
\ee
which connects asymmetric minima. 

\begin{figure}[h!] 
\includegraphics[scale=0.45]{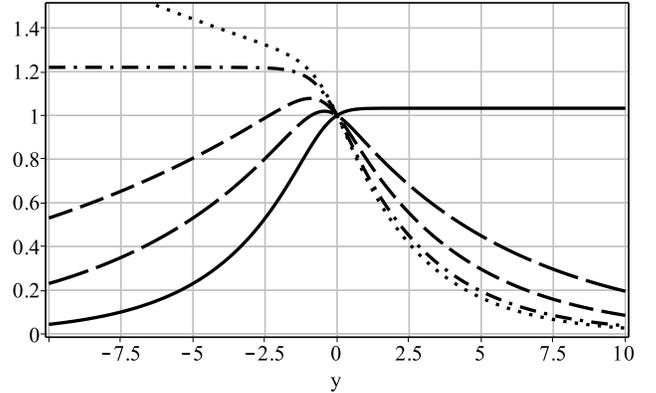} 
\caption{The warp factor $e^{2A}$ with $A$ given by Eq. \eqref{warp6} for $c=-1/2$ (solid line), $c=-1/4$ (long dashed line), $c=-1/8$ (dashed line), $c=0$ (dot-dashed line), and $c=1/20$ (dashed line).}\label{fig5}
\end{figure}

\begin{figure}[h!] 
\includegraphics[scale=0.45]{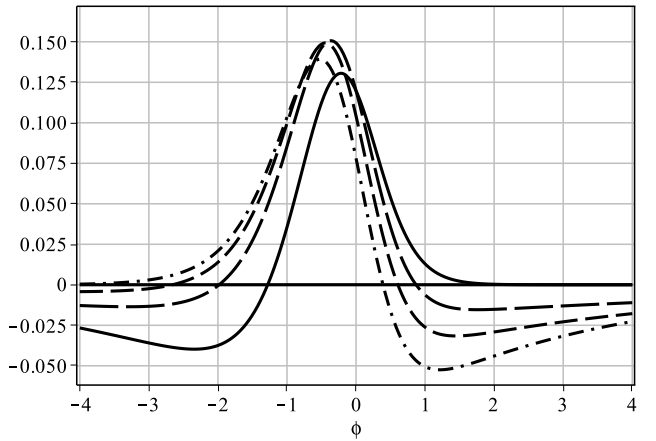} 
\caption{The energy density given by Eq. \eqref{rho6} for $c=-1/2$ (solid line), $c=-1/4$ (long dashed line), $c=-1/8$ (dashed line), $c=0$ (dot-dashed line), $c=1/20$ (dotted line).}\label{fig6}
\end{figure}

Here the warp function is given by
\be\label{warp6}
A(y)=-\frac1{24} \tanh(y) + \frac1{12} \ln(\sech(y)) - \left(\frac1{12}+\frac{c}{3}\right)y\,,
\ee
and presents the asymptotic values
\bes
\ben
A_{\pm \infty}(y)&=&-\frac13\left[\frac1{4}\pm \left(\frac1{4}+{c}\right) \right]|y|\,.
\een
The bulk cosmological constant is provided by 
\ben
\Lambda_{5_\pm}&=&-\frac13\left[\frac1{4}\pm \left(\frac1{4}+{c}\right) \right]^2.
\een
\ees
For $c<-1/2$ and $c>0$, the warp factor diverges. We obtain the asymmetric branes, separating two bulk spaces $\mathbb{M}_5$ and A$dS_5$, for $c=-1/2$ and $c=0$. In the first case, $\Lambda_{5+}=0$ and $\Lambda_{5-}=-1/12$, while in the  second  case $\Lambda_{5+}=-1/12$ and $\Lambda_{5-}=0$.
The energy density is 
\ben\label{rho6}
\rho(y)&=&e^{2A} \Bigg[\frac{39}{64}-\frac{c}{12}-\frac{c^2}{9} - \left(\frac{31}{48}-\frac{c}{18}\right)\tanh(y) \nonumber\\
&&+\left(\frac{c}{36}-\frac{35}{288}\right)\tanh^2(y)+\frac{17}{144}\tanh^3(y)\nonumber\\&&-\frac{1}{576}\tanh^4(y)\Bigg].
\een
In Figs.~\ref{fig5} and \ref{fig6} we depict the warp factor and the energy density, for some values of $c$.
%%%%%%%%%%%%%%%%%%%%%%%%%%%%%%
%%%%%%%%%%%%%%%%%%%%%%%%%%%%%%

%%%%%%%%%%%%%%%%%%%%%%%%%%%%%%
%%%%%%%%%%%%%%%%%%%%%%%%%%%%%

This model is different from the previous one. The potential is always symmetric, and the kinklike solution connects asymmetric minima. The model gives rise to warp factor that is always asymmetric, although at $c=-1/4$ it connects same $AdS_5$ bulk spaces.

%%%%%%%%%%%%%%%%%%%%%%%%%%%%%%%%
\section{Stability}

An issue of interest concerns stability of the gravity sector of the braneworld model.
For the models studied in the previous section, such investigations can be implemented
numerically \cite{bgl}, an issue to be described in the longer work under preparation \cite{blmr}. 
Here, however, to gain confidence on the stability of the suggested braneworld scenarios,
we develop the analytical procedure: we get inspiration from the first work in Ref.~\cite{G}, where it is
shown that the sine-Gordon model is good model for analytical investigation. Thus, we consider
the sine-Gordon-type model with 
\ben
W_c(\phi) = 2\sqrt{\frac{3}{2}}\sin\left(\sqrt{\frac{3}{2}}\phi\right)+c,\label{wp}
\een\noindent
with $|c|\leq\sqrt{6}$. We note that for $\phi$ small, the model is similar to the case of an odd superpotential studied previously;
thus, the current investigation engenders results that are also valid for the model investigated in Sec.~\ref{sec:2a}. The point here is
that we want to study stability analytically, so the sine-Gordon model is the appropriate model.

We use Eqs.~(\ref{foA}) and (\ref{firstorderA}) to obtain analytic solutions
\bes\ben
\phi(y)&=& \sqrt{\frac{3}{2}}\arcsin(\tanh(y)),
\\
A_c(y)&=& A_0(y)-\frac13 c y,
\\
A_0(y)&=&-\ln(\sech(y))\,,
\een\ees 
where $\phi(y)$ and $A_0(y)$ are the standard solutions of the sine-Gordon model, for $c=0$.

To study stability, we follow Ref.~\cite{bgl}. We work in the tranverse traceless gauge,
to decouple gravity form the scalar field. Also, we have to change from $y$ to the conformal coordinate $z=z(y)$
to get to a Schroedinger-like equation with a quantum mechanical potential. This investigation cannot be done analytically anymore,
unless we take $c=0$. Thus, we resort on the approximation, taking $c$ very small and expanding the results up to first-order in $c$.
In this case we can write the conformal coordinate as
\bes\ben
 z(y)&=& f(y) +\frac{c}3 \left(y f(y) - \int dy f(y)\right),\label{zzz}
\\
f(y)&=&\int dy \, e^{-A(y)}.
\een\ees 
We note that $f(y)$ is an even function, while the $c$-term contribution is odd. Therefore $z(y)$ is an asymmetric function. 
The inverse is
\be
y=f^{-1}(z)-\frac{c}{3}\left[\frac{1}{f^\prime(x)}\left(xz\!-\!\!\int dx f(x)\right)\right]_{x=f^{-1}(z)}.
\ee
After some algebraic calculations, we could find the associated potential
\be
U(z)=\frac32 A^{\prime\prime}+\frac94 A^{\prime2}+c \,U_c(z)\label{alinha}
\ee
where $U_c(z)$ is the contribution up to first-order in $c$. 
The sine-Gordon-type model is nice, and we could find the explicit results:
the conformal coordinate $z=\int e^{-A(y)}\,dy$ in (\ref{zzz}) is computed as 
\ben
\!\!\!\!\!\!\!\!z(y)&=& \sinh(y) + \frac{c}{3}\left(y\,\sinh(y)-\cosh(y)\right)\,.
\een
Since $c$ is small, the above expression can be inverted to give
\ben
y(z)&=& \arcsin\!{{\rm h}}(z)-\frac{c}{3}\!\left[\!\frac{1}{\sqrt{1+(z)^2}}\right.\nonumber\\&&\left.\!\times\left(\;(z) \arcsin\!{\rm h}(z)- \sqrt{1+(z)^2}\right)\right].
\een\noindent
Therefore 
\ben\label{Az}
A(z)&=& -\frac{1}{2}\ln\left(1+(z)^2\right)\nonumber\\&&-{c}\left(\frac{\arcsin\!{\rm h}(z)}{1+(z)^2}+\frac{z}{\sqrt{1+(z)^2}}\right).
\een\noindent

%%%%%%%%%%%%%%%%%%%%%%%%%%%%%%%%%%
\begin{figure}
\includegraphics[scale=0.45]{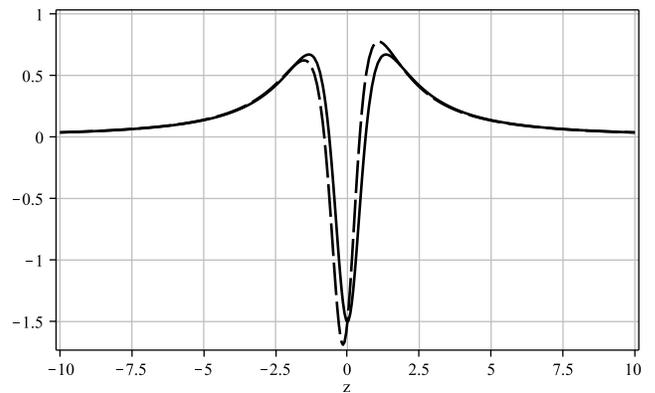} 
\caption{\label{fig9}Plot of the quantum mechanical potential in (\ref{qmp}), for $c=0$ (solid line), $c=1/5$ (long dashed line).}
\label{fig7}
\end{figure}
%%%%%%%%%%%%%%%%%%%%%%%%%%

It follows from Eq.(\ref{alinha}) that the quantum potential is written as 
\bes\label{qmp}\ben
U(z)&=& \frac{3}{4}\frac{(5z^2-2)}{(1+z^2)^2}+c\,U_c(z)\label{uz}
\\
U_c(z)&=&\frac{1}{(1+z^2)^3}\left((1-6 z^2)\arcsin\!{\rm h}(z)+\right.\nonumber
\\&&\left.7z\sqrt{1+z^2)}\right)\label{uzc}
\een\ees
It is depicted in Fig.~\ref{fig7}. The maxima of the potential ${U}(z)$ are provided by
\ben
z_\pm &=&\pm\frac{3}{\sqrt{5}}-\frac{\sqrt{70}c}{9450}\left(27 \sqrt{5} \arcsin\!{\rm h}\left(\frac{3}{\sqrt{5}}\right)\sqrt{70}-1330\right) \nonumber\\&\approx& \pm1.34164-0.68398\,c
\een
In the limit $z\to\infty$, the asymptotic value of the potential reads
\ben
{U}(z) \approx \frac{15}{4z^2} -\frac{1}{z^4}(9+c(6\ln(2z)-7))+\mathcal{O}\left(\frac{1}{z^6}\right).
\een
It explicitly shows how the asymmetric contribution enters the game asymptotically.

\begin{figure} 
\includegraphics[scale=0.45]{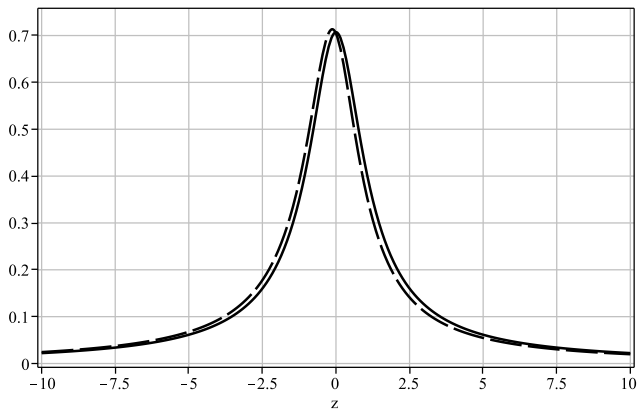} 
\caption{\label{fig9}Plot of the normalized graviton zero mode $\psi_0(z)$, for $c=0$ (solid line) and for $c=1/5$ (long dashed line).}
\label{fig8}
\end{figure}

The quantum mechanical analogous problem is described by 
the equation
\be\left(-\frac{d^2}{dz^2}+{U}(z)\right)\psi_n=k^2\psi_n\,.
\ee
This Hamiltonian can be fatorized as, up to first-ordem in $c$,
\be
-\frac{d^2}{dz^2}+U(z)=\left(-\frac{d}{dz}+\frac32 A^\prime\right)\left(\frac{d}{dz}+\frac32 A^\prime\right)
\ee
where $A=A(z)$ is given by \eqref{Az}. It is then non-negative, so there are no negative bound states. In fact, the graviton zero mode is $\psi_0(z)$, which solves the above equation for $k=0$; it is given explicitly by 
\ben
\psi_0(z) &=& \exp\left(\frac{3}{2}A(z)\right)\nonumber\\
&=&\left(1+(z)^2\right)^{-\frac{3}{4}}\times\nonumber\\&&\!\!\!\!\left(1-\frac{c}{3}\left(\frac{\arcsin\!{\rm h}(z)}{1+z^2}+\frac{z}{\sqrt{1+z^2}}\right)\right),
\een
and is depicted in Fig.~\ref{fig8}, showing its asymmetric behavior. It has no node, indicating that it is the lowest bound state, as expected.
%%%%%%%%%%%%%%%%%%%%%%%%%%%%%%%%%%
%%%%%%%%%%%%%%%%%%%%%%%%%%%%%%%%

%%%%%%%%%%%%%%%%%%%%%%%%%
\section{Concluding Remarks}

In this work we investigated the presence of asymmetric branes, constructed from the addition of a constant $c$ to $W(\phi)$, the function that defines the potential
in the form$$
V(\phi)=\frac18\,W^2_\phi-\frac13\,W^2.
$$
This expression controls the scalar field model that generates the thick brane scenario, and it is important to reduce the equations to its first-order form
\be
\phi^\prime=\frac12 W_\phi;\;\;\;\;\;A^\prime=-\frac13 W.
\ee

 The main results of the current work show that we can construct asymmetric brane, irrespective of the scalar potential
being symmetric or asymmetric.

We have investigated two distinct models, of the $\phi^4$ and $\phi^6$ type, and we have explicitly shown how the constant added to $W(\phi)$ works to induce asymmetry
to the thick braneworld scenario. Moreover, we also studied the sine-Gordon type model, focusing on the stability of the gravitational scanario. The nice properties of the sine-Gordon model
very much helped us to implement analytical calculations, showing stability of the graviton zero mode, despite the presence of an asymmetric volcano potential.

Evidently, the asymmetry induced in the thick braneworld scenario may be phenomenologically relevant, since it has to be consistent with the $AdS_5$ bulk curvature, and to the experimental and theoretical limits for the brane thickness \cite{kapp,epl}.
 
When finishing this study, we become aware of the work \cite{aa}, which investigates similar possibilities.

The authors would like to thank CAPES and CNPq for partial financial support.

%%%%%%%%%%%%%%%%%%%%%%%%%%%%%%%%%

\end{document}